\documentstyle[12pt,epsf]{article}
\begin{document}
\newcommand{\fm}{ferromagnet}
\newcommand{\tm}{transition metal}
\newcommand{\be}{\begin{equation}}
\newcommand{\ee}{\end{equation}}
\newcommand{\bea}{\begin{eqnarray}}
\newcommand{\eea}{\end{eqnarray}}

\title{Spin and Orbital Order in Itinerant Ferromagnets}

\author{P. Fazekas\\
{\small Research Institute for Solid State Physics and Optics}\\
{\small Budapest 114, P.O.B. 49, H-1525 Hungary }}

\date{}

\maketitle

\begin{abstract}
The long-standing problem of the effect of correlations on the ferromagnetism 
of {\tm}s is apparently nearing solution. The {\fm}ism of {\tm} compounds, for
instance doped manganites, poses a new question: is there some kind of orbital
order coexisting with itinerant {\fm}ism? The ideas and techniques introduced
by Gutzwiller should be of use again. 
\end{abstract}

\section{Introduction}

In the years 1963 to 1965, Gutzwiller published three highly influential
papers \cite{Gutz1,Gutz2,Gutz3} about the ``Effect of Correlation on the 
Ferromagnetism of Transition Metals''. The problem had been long considered 
before but it came to a deadlock: mean field criteria (Stoner theory) gave 
{\fm}ism all too easily. Since Van Vleck's classic analysis \cite{VV} it was
also known what the difficulty is: Stoner theory neglects the correlation 
between opposite-spin electrons, and therefore overestimates the kinetic 
energy gain which arises from making the spins parallel. One needed a method
for describing local correlations between itinerant electrons. Clearly, the 
problem is ubiquitious in electronic systems, and has many manifestations
beside the onset of {\fm}ism. If one can say why Fe, Co, and Ni are 
{\fm}ic while the corresponding $4d$ elements Ru, Rh, and Pd are not, one 
can probably also say why CoO is
insulating - perhaps even more easily. And so it was to turn out. The
threefold contribution of Gutzwiller: a hamiltonian, a variational method, and
an approximate analytical solution, proved to be of universal value; and
of all the related problems, the {\fm}ism of the transition metals was one of
the most difficult. 

In the next Section, I briefly describe why the problem of {\fm}ism is
particularly difficult, and what our present view is. Though it does not come 
into this paper, it may be mentioned that the puzzle of the high-$T_C$ 
superconductivity of cuprates inspired a keen re-examination of all
correlation problems, and thus indirectly contributed to a renaissance of the
subject pioneered by Gutzwiller. Lately, we have been witnessing the 
important development that the 
combination of the Gutzwiller variational approach with LDA band structure 
calculation promises to become one of the major techniques of computational 
solid state physics. It has been demonstrated that starting with nine 
($s$, $p$, $d$) orbitals per site, and using a judiciously chosen number 
of variational parameters to describe the possible on-site spin--spin, 
orbital--spin and orbital--orbital correlations, an impressive array of 
the electrical, magnetic and optical properties of nickel can be 
calculated, with an overall quality which surpasses that of spin density 
functional theory \cite{flori}. One might perhaps claim that the program 
initiated by Gutzwiller has been brought to a successful conclusion, 
using the very tools he devised. The attractive feature of the approach
outlined in \cite{flori} would be that 
(in spite of the inevitable complexity of the procedure) one still has a 
clear idea which correlations have been taken into account and what 
their role is.

Even within the family of $3d$ systems, the nature of {\fm}s is quite varied
 and while essentially full undestanding may be in sight for some of them,
 basic questions remain to be answered for others. The {\fm}ic
transition metal compound La$_{1-x}$Sr$_x$MnO$_3$ comes to mind 
\cite{scien2000}. One may assume that if we understand the {\fm}ism of Ni, 
we should understand that of La$_{1-x}$Sr$_x$MnO$_3$ even better, since 
this compound gives a better realization of a Gutzwiller--Kanamori--Hubbard 
model with partial band filling than Ni does: it does not have the 
complication of an overlapping $4s$ band. However, the nature of {\fm}ic 
manganites is a matter of lively debate \cite{khom2000,shiba2000,naga2000}. 
Though the problem of complex orbital order (Sec. 3) does not have the same
broad significance as ordinary ferro- or anti{\fm}ism, it has some 
amusing parallels with the long-studied problem of spin ordering, and leads to
 a  series of questions which are not unlike those which Gutzwiller set out to 
investigate long ago.

\section{Ferromagnetism of Correlated Electrons}

In \cite{Gutz1} and \cite{Gutz2}, Gutzwiller discussed the key 
ingredients which give rise to the {\fm}ism of transition metals, particularly
nickel. He distingushed between conduction ($4s$),  and the almost 
localized `valence' ($3d$) electrons, whereby the interactions among the 
latter kind of electrons were identified as the cause of magnetism. Though his 
description in words amounts to a recipe for the periodic Anderson model
(including hybridization), Gutzwiller decided to omit the conduction
electrons, and kept only the band of interacting electrons in the
Hamiltonian\footnote{Nowadays we refer to it as the Hubbard model, but it was
  introduced  independently by Gutzwiller \cite{Gutz1}, 
Hubbard \cite{Hubb}, and Kanamori \cite{Kana} in the same year.}. First, 
let us write down the 1-band version
\be
{\cal H}_{\rm 1-band} = -t\sum_{\langle {\bf i},{\bf j}\rangle} \sum_{\sigma}
(c_{{\bf i}\sigma}^{\dagger}c_{{\bf j}\sigma} + {\rm H.c.}) + 
U\sum_{\bf j} {\hat n}_{{\bf j}\uparrow}{\hat n}_{{\bf j}\downarrow}\, . 
\label{eq:1}
\ee
The dichotomy of itinerant and localized features (embodied in the hopping
and on-site interaction terms, resp.) makes the analysis of this simple-looking
hamiltonian extremely difficult. Being aware of the pitfalls of a 
mean-field-like smearing-out of the interaction term, Gutzwiller insisted that
the many-electron state should contain local correlations. This is achieved by
gradually suppressing the double occupation of sites in a spin-polarized 
Fermi sea: 
\be
|\Psi\rangle = \prod_{\bf j}[1-(1-\eta){\hat n}_{{\bf j}\uparrow}
{\hat n}_{{\bf j}\downarrow}]\,
\prod_{\bf k}^{|k|<k_{{\rm F}\uparrow}}c_{{\bf k}\uparrow}^{\dagger} 
\prod_{\bf k}^{|k|<k_{{\rm F}\downarrow}}
c_{{\bf k}\downarrow}^{\dagger}|0\rangle
\label{eq:2}
\ee
There are two variational parameters: $\eta$ which controls the local
correlations, and the polarization density $m=(N_{\uparrow}-N_{\downarrow})/L=
n_{\uparrow}-n_{\downarrow}$, where $L$ is the number of lattice sites. $\Psi$
can be used for any band filling $n=n_{\uparrow}+n_{\downarrow}$. In spite of
its apparent simplicity, $\Psi$ is very difficult to handle: in order to
evalute $n_d$, the density of doubly occupied sites, the Fermi sea has to be
decomposed into $\sim e^L$ localized configurations. The full implementation
of the variational procedure became possible only 24 years later \cite{MV2}. 
However, a great variety of approximate variational solutions came earlier,
thanks to a technical innovation announced in Gutzwiller's third paper 
\cite{Gutz3}: the Gutzwiller approximation\footnote{Ref. \cite{LN} gives 
an introduction to the Gutzwiller variational method.} (in short GA). 
It proposes 
a practicable analytical method for solving the many-electron variational 
problem formulated in the letter \cite{Gutz1}. The essential result is that
the kinetic energies get scaled down from their $U=0$ values by the
spin-dependent renormalization factors $q_{\sigma}$, and the energy expression
becomes 
\be
\left.\frac{\langle\Psi|{\cal H}_{\rm 1-band}|\Psi\rangle}
{\langle\Psi|\Psi\rangle}\right|_{\rm GA} = \sum_{\sigma}q_{\sigma}
{\bar\epsilon}_{\sigma} + Un_{\rm d}
\label{eq:3}
\ee
Here $q_{\sigma}$ still depends on $n_{\rm d}$ which has taken over the role
of the original variational parameter $\eta$. It is enlightening to quote the 
strong coupling limiting form of $q_{\sigma}$
\be
q_{\sigma}(U\to\infty) = \frac{1-n}{1-n_{\sigma}}\, .
\label{eq:4}
\ee
For a fully polarized system $n_{\uparrow}=n$, $q_{\uparrow}=1$, the kinetic
energy keeps its bare value. This echoes the old exchange hole argument for
the arising of {\fm}ism: parallel-spin electrons can avoid each other, there
is no interaction energy penalty, but it comes at a price: the majority spin
band has to be filled up to a higher Fermi energy. In contrast, the 
paramagnetic solution
keeps the Fermi level deeper down in the band, but the kinetic energy gain is 
reduced by correlation-induced band narrowing:
\be
q_{\uparrow}(U\to\infty) =q_{\downarrow}(U\to\infty) =\frac{1-n}{1-(n/2)}<1\, .
\label{eq:5}
\ee
For bands with a symmetrical density of states, the {\fm}ic and 
paramagnetic solutions 
reach a balance at about quarter filling, and the low-density system remains 
non-magnetic even at $U\to\infty$ \cite{FMM90}. This should be contrasted with 
the Stoner result that the system becomes {\fm}ic at arbitrary filling if $U$
is large enough. The criteria for the onset of antiferromagnetism are also
strongly modified \cite{MV1,FMM90}.

In fact, at half-filling ($n=1$), $q(U\to\infty)\propto (1-n)=0$, signalling 
 that we are within the Mott insulating phase.  Being primarily interested 
in itinerant {\fm}ism, Gutzwiller did not pay attention to the particular 
case of integral band filling but it was soon realized by Brinkman and Rice 
\cite{BR70} that a critical divergence of the effective mass of the electrons 
on the metallic side of the Mott transition is a corollary of the
results presented in \cite{Gutz3}. The increasing ``heaviness'' of the
 fermions is  manifested in an enhancement of the electronic specific heat 
and the spin susceptibility, and a reasonable Wilson ratio can be derived. 
This was the beginning of our understanding of the $T=0$ Mott transition 
as a quantum phase transition; up-to-date studies using dynamical mean 
field theory still yield results which bear remarkable  similarity to 
the Gutzwiller--Brinkman--Rice scenario (see, e.g. \cite{schlipf}). 

Subsequent studies of the periodic Anderson model (in which the strongly 
correlated band has non-integral filling) suggested a similar interpretation 
of the heavy Fermi sea of almost-integral-valent $f$-electron systems 
\cite{RU86,FB87}.  For a wide range of problems, the Gutzwiller method 
allowed to develop a pictorial way of thinking about local correlation 
effects, and also provided the technique to do the corresponding 
calculations. The generalization of the ideas to degenerate bands, and to 
several correlated bands, takes a lot of work but is 
conceptually straightforward \cite{bune,itai}. The use of the variational
method is not constrained by the GA; we now understand that GA is really an
infinite-dimension ($D=\infty$) approximation, and we can go beyond it by
either exact evaluation \cite{MV2}, or by calculating 
$1/D$-corrections for $D>2$ \cite{Geb91}. 

Though the Gutzwiller method goes a long way to correct the Hartree--Fock 
results \cite{penn} for the magnetic phase diagram, we have reason to 
suspect that it still overestimates the chances of {\fm}ic ordering. 
It is a typical Gutzwiller result that the extent of the FM phase along the
$n$ axis is largest at $U\to\infty$. This justifies another look at the
mechanism driving {\fm}ism in the strong coupling limit. Performing 
the canonical transformation which eliminates double occupation 
in lowest order, we arrive at the familiar $t$--$J$ model:
 \bea
{\cal H}_{t-J} & = & -t\sum_{\langle i,j\rangle}\sum_{\sigma}(
{\tilde c}_{i\sigma}^{\dagger}{\tilde c}_{j\sigma} + {\rm H.c.}) + 
\frac{4t^2}{U}\sum_{\langle i,j\rangle}\left({\bf S}_i{\cdot}{\bf S}_j-
\frac{{\hat n}_{i}{\hat n}_{j}}{4}\right)\nonumber\\[2mm]
& &  + ({\rm three-site\ \ terms})
\label{eq:tJ}
\eea
Let us note that the 1-band model generates only one kind of effective 
spin--spin interaction and that is {\sl anti{\fm}ic}. This need not rule out
the possibility of a {\fm}ic ground state  but if it exists, the reason 
for it is not straightforward. Given that itinerant {\fm}ism 
(if any) does not originate from the interaction term in (\ref{eq:tJ}), 
it has to arise from the projected kinetic energy; it is an elusive, and 
often fragile, phenomenon. 
Since the nature of a uniformly spin-polarized state is so easy to grasp, 
it is somewhat surprising to find out 
that the {\fm}ism of the 1-band model poses a much more difficult many-body
problem than anti{\fm}ism, and that even for three-dimensional systems, mean
field predictions are grossly misleading \cite{wahle}. After thirty years of
continuing efforts we know that 
Gutzwiller's  immediate reaction \cite{Gutz1}: ``It seems rather that the 
exact ground state of the Hamiltonian (1) is never {\fm}ic'' was not 
completely right but it captured the essential point that the existence 
of a robust {\fm}ic phase is not a generic property of the 1-band model. 
For bipartite lattices with a symmetrical density of states, 
even if a {\fm}ic phase exists, it is only at such large coupling strengths 
that we can safely forget about it in connection with real systems. 
 However, intermediate-coupling {\fm}ism was shown to
appear for non-bipartite lattices (e.g. fcc) or/and a non-symmetric 
band density of states \cite{wahle}. The relevance of these factors was
already pointed out in the pioneering works of Gutzwiller, Kanamori, and 
Hubbard; Gutzwiller also realized that direct exchange which is
ordinarily negligible compared to other Coulomb terms, can become important 
when the magnetic and non-magnetic states are finely balanced.

\section{Orbital Liquid vs Complex Orbital Order}

Looking back, we might say that the detour via the single-orbital model
(\ref{eq:1}) made the understanding of {\fm}ism unnecessarily difficult. In
his first papers, Gutzwiller clearly stated that the degeneracy of the
$d$-band, and the intra-shell Hund exchange, are of vital importance. However,
technical difficulties forced us to start with the 1-band model. This has
brought many important results (foremost an understanding of the Mott
transition), but it gradually became clear that the model is very reluctant to
yield itinerant {\fm}ism at reasonable couplings. We now largely understand
why this is the case, but perhaps we should never have worried too much: most
of the itinerant {\fm}s we are interested in are based on degenerate
shells. Models with orbital degeneracy have several efficient mechanisms to
give spin {\fm}ism. However, they bring the additional complications of
orbital ordering or/and orbital fluctuations.

A model with twofold orbital degeneracy is relevant for the doped {\fm}ic 
manganites such as La$_{1-x}$Sr$_x$MnO$_3$ which are Mn$^{3+}$--Mn$^{4+}$ 
valence fluctuators. Both Mn$^{3+}$ and Mn$^{4+}$ have stable $T_2^3$ cores 
in the $S=3/2$ high-spin state. The metallic nature arises from the fourth
electron of Mn$^{3+}$ ions which may hop around in the $E$-shell states. For a
realistic description of manganites, it is important to remember the $T_2^3$ 
cores, and the $T_2$--$E$ intra-atomic exchange \cite{heldvol,naga2000}, but 
the basic questions of spin magnetism and orbital order can be discussed
within the 2-band model. We write it down for the pair of sites 1, 2, which is
sufficient for deriving the strong-coupling effective hamiltonian. $a$ and $b$
denote the orbitals. In general, both the interorbital and intraorbital
hopping amplitudes differ from zero 
\bea
{\cal H}_{\rm hop} & = & -t_a\sum_{\sigma}(c_{1a\sigma}^{\dagger}c_{2a\sigma} 
+ {\rm H.c.}) -t_b\sum_{\sigma}(c_{1b\sigma}^{\dagger}c_{2b\sigma} + 
{\rm H.c.})\nonumber\\
& & -t_{ab}\sum_{\sigma}(c_{1b\sigma}^{\dagger}c_{2a\sigma} +
c_{1a\sigma}^{\dagger}c_{2b\sigma} + 
{\rm H.c.})\, .
\label{eq:orbh1}
\eea

We do not write down the familiar interaction terms which contain the intra-
and interorbital Hubbard terms (with $U_a$, $U_b$, and $U_{ab}$), and the 
{\fm}ic Hund coupling $-2J$. Actually taking the cubic $E$ doublet would
require a specific relationship between the parameters, but we keep the
notation general because we intend to apply it to the discussion of other
crystal field levels as well.

The $2\times 2$ (spin$\times$orbital) degeneracy of the shell states is
conventionally expressed by representing the orbitals as $\tau=1/2$ pseudospin
states: $\varphi_a\rightarrow |\tau^z=+(1/2)\rangle$ and 
$\varphi_b\rightarrow |\tau^z=-(1/2)\rangle$. This allows to write the
large-$U$ effective hamiltonian as mixed spin--pseudospin interaction. Its
derivation from the 2-band Hubbard model is straightforward but tedious. We do
not quote the full result (the familiar Kugel-Khomskii model 
\cite{Kukh}), merely the term which acts on spin triplets:
\bea
{\cal H}_{\rm eff}^{\rm tr} & = & \frac{1}{U_{ab}-2J}
\left(\frac{3}{4}+{\bf S}_1{\cdot}{\bf S}_2\right)\left\{ {\rm constant}+
4(t_at_b-t_{ab}^2)\,{\mbox{\boldmath{$\tau$}}}_1{\cdot}
{\mbox{\boldmath{$\tau$}}}_2 \right.\nonumber\\
& & + 
\left. 2(t_a-t_b)^2\tau^z_1\tau^z_2 + 8t_{ab}^2 \tau^x_1\tau^x_2 + 
4t_{ab}(t_a-t_b)(\tau^x_1\tau^z_2 + \tau^z_1\tau^x_2)\right\} 
\label{eq:effham}
\eea
 The spin-singlet term is similar. Let us observe that in general, the 
interaction is anisotropic in pseudospin space. 

\begin{figure}[ht]
\epsfxsize=13.0cm
\begin{displaymath}
\centerline{\epsfbox{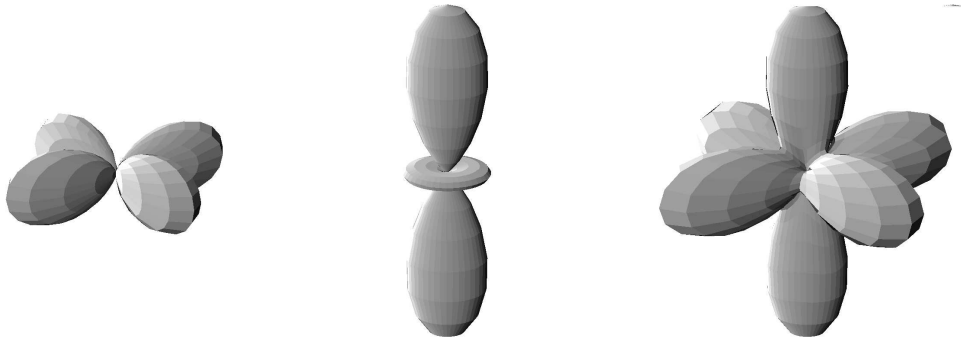}}
\end{displaymath}
\caption{\small Left and middle: the conventional real basis functions 
$\varphi_a\sim x^2-y^2$ and $\varphi_b\sim 3z^2-r^2$ for the cubic $E$ 
doublet. Right: the complex combinations $\varphi_a\pm i\varphi_b$ are 
octupolar eigenstates.\label{fig:cub}}
\end{figure}

The usual choice of basis for the cubic $E$ doublet is shown in
Fig. \ref{fig:cub}. The real basis states have lobes in specific directions. 
The hopping amplitudes in (\ref{eq:orbh1}) are not merely orbital index
dependent, but depend also on the direction of the hopping.

Let us now look for the {\fm}ic ground state of the two-band model. For the
sake of simplicity, we may assume full spin polarization, so that we once
again have only two states per site (the two orbitals), like in the 
simple 1-band model (\ref{eq:1}). It may seem that the remaining problem of an 
{\sl orbitally  correlated} state  is at the same level of difficulty as 
the thoroughly studied correlation problem of the ordinary Hubbard model 
\cite{rozenberg}. However, the orbital index dependence of the hopping 
matrix elements brings additional complications: it is as if in the 
single-orbital model, we had spin-dependent hopping, with spatial and spin 
anisotropy. In the case of (\ref{eq:1}), if we find a spin-{\fm}ic state, 
we know that the total spin can point in any direction. For the 
two-band model, polarization in the pseudospin $z$-direction is something 
totally different from polarization along the $y$-direction.

The ground state we seek should be analogous to (\ref{eq:2}) but how to
specify the orbitals? Postulating an orbital order of the kind found in 
insulating LaMnO$_3$ would mean that each unit cell has an asymmetric 
form of the $3d$ electron cloud, and an accompanying Jahn--Teller 
distortion. However, experiments tell us that in the {\fm}ic metal the
Jahn--Teller distortion is absent\footnote{At least in some of the
  manganites.} and the unit cells should be thought 
of as cubic. An obvious idea is that 
this is the result of time-averaging: the ground state is 
orbitally disordered (``pseudospin paramagnetic''). Let us take the
non-interacting two-band Fermi sea and Gutzwiller-project it 
(now two electrons can share the same site only if they are in different 
orbitals)
\be
|\Phi\rangle = \prod_{\bf j}[1-(1-\eta){\hat n}_{{\bf j}a\uparrow}
{\hat n}_{{\bf j}b\uparrow}]\,
\prod_{\bf k}^{|k|<k_{{\rm F}\alpha}}c_{{\bf k}\alpha\uparrow}^{\dagger} 
\prod_{\bf k}^{|k|<k_{{\rm F}\beta}}
c_{{\bf k}\beta\uparrow}^{\dagger}|0\rangle
\label{eq:22}
\ee 
where $\alpha$, $\beta$ refer to the two bands (the band structure is shown in 
\cite{shiba2000}). In spite of orbital disorder, the ground state is unique 
(an orbital liquid \cite{naga2000}), in the same sense that the ordinary Fermi
sea is a spin liquid.  

Is there some other possibility? Can we prescribe some kind of orbital order
and still get a similar picture of a cubic ferromagnetic metal? Obviously, it
could not be done by ascribing  one of the real wave functions
$|x^2-y^2\rangle$ or $|3z^2-r^2\rangle$ to the sites. However, there is another
possibility. The order parameter space is described by the product
representation $E\otimes E=A_1+A_2+E$. $E$ belongs to the familiar 
quadrupolar doublet of $(x^2-y^2)$-like and $(3z^2-r^2)$-like order parameters,
but we have not yet considered the single component order parameter with 
$A_2$ symmetry. Since it transforms like $xyz$, it must be the magnetic
octupolar moment ${\overline{L_xL_yL_z}}$ \cite{shiba2000} where the bar 
indicates symmetrization over six terms 
\be
{\overline{L_xL_yL_z}}\rightarrow \frac{1}{6}(L_xL_yL_z+L_xL_zL_y+\dots)=
\frac{i}{4}(L^+L^zL^+-L^-L^zL^-)\propto \tau^y .
\label{eq:tauy1}
\ee
The $\tau^y=\pm(1/2)$ eigenstates are the complex combinations 
\be
|\zeta_{\pm}\rangle=\frac{1}{\sqrt{2}}\left(|x^2-y^2\rangle 
\pm i|3z^2-r^2\rangle \right)
\label{eq:tauy2}
\ee
First of all, let us observe that the charge density contours of these complex
orbitals show cubic symmetry (Fig. \ref{fig:cub}, right). Postulating the 
uniform octupolar state
\be
|\Omega_0\rangle=\prod_{\bf j} \left(c_{{\bf j}a\uparrow}^{\dagger} + 
ic_{{\bf j}b\uparrow}^{\dagger}\right) |0\rangle
 \label{eq:tauy3}
\ee
it would show the cubic symmetry demanded from the ferromagnet. Naturally, 
$|\Omega_0\rangle$ would be a fully localized state with one electron at 
each site, so it is not yet what we want. To have a uniformly octupolar 
itinerant spin ferromagnet, we have to construct a trial state analogous 
to (\ref{eq:2}), a Gutzwiller-projected Fermi sea with a finite octupolar 
polarization \cite{pfi}.

The arising of complex order as a possibility distinct from real order follows
from the lack of rotational symmetry in isospin space. There is another
interesting aspect: the possibility of time reversal invariance breaking
without (the usual kind of) magnetism. Actually, it occurs whenever we have an
orbital (non-Kramers) doublet to start with. The local basis can always be
chosen in a complex form, and there will be a corresponding complex operator 
to which these are eigenstates (in the present case, it is the purely
imaginary $\tau^y$). But a complex quantity changes to its conjugate under
time reversal, thus the state is breaking that symmetry.

In which form time reversal invariance breaking happens, depends on the kind
of doublet we start from. In the present case, it is broken by a magnetic
octupolar component, which has been reintroduced into solid state physics only
recently, mainly with reference to CeB$_6$ \cite{sakai,kura}. It is, however,
interesting to recall that the very  problem we are discussing, namely 
the ordering patterns supported by cubic $E$ subshells, has been considered in
the early work by Korovin and Kudinov \cite{KK}. These authors remarked that
postulating a state like $|\Omega_0\rangle$ ``...leads to nonzero values 
of the third moments 
$\int x_{\alpha}x_{\beta}x_{\gamma}j_{\delta}({\bf r})d{\bf r}$ of the current
density $j({\bf r})$ in the crystal\footnote{The macroscopic magnetic
  octupolar moment introduced by Korovin and Kudinov is not literally the same
  as the order parameter (\ref{eq:tauy1}) considered in \cite{shiba2000}, but
  the two concepts are related.} (the first moment, i.e., the usual
orbital magnetic moment, is absent for the $E_g$ representation)''. The
suggestion was then apparently forgotten because insulating LaMnO$_3$ has real
orbital order (it is Jahn--Teller distorted), and the possibility of an
essentially invisible orbital order in the cubic {\fm}ic phase was not
taken seriously. However, the resurgence of general interest in higher
multipolar order, and the availability of sophisticated experimental
techniques for its detection,  led to a reconsideration of 
the octupolar phase \cite{khom2000,shiba2000}. In particular, 
Takahashi and Shiba \cite{shiba2000} determined the orbital analogue of the
Penn phase diagram \cite{penn}, comparing the threshold values for the RPA 
instabilities against various kinds of uniform and staggered orders, and found
that staggered octupolar order is the first instability in a wide range of
band filling. By its nature, theirs is a weak coupling calculation, and it is
clearly desirable to complement it by a strong-coupling, Gutzwiller-like
calculation.    

One reason why octupolar order may have long eluded consideration, is that it
is not motivated by the strong coupling effective hamiltonian
(\ref{eq:effham}). For the cubic $E$ doublet the hopping amplitudes 
in the $x$--$y$ plane obey 
$t_a:t_b:|t_{ab}|=3:1:\sqrt{3}$, thus the coefficient of 
$\tau_1^y\tau_2^y$ vanishes. It is a little bit like the story of 
{\fm}ism in the 1-band model
where we did not have {\fm}ic spin--spin interaction in the $t$--$J$ model,
but nonetheless felt justified in looking for a {\fm}ic ground state.

\begin{figure}[ht]
\epsfxsize=13.0cm
\begin{displaymath}
\centerline{\epsfbox{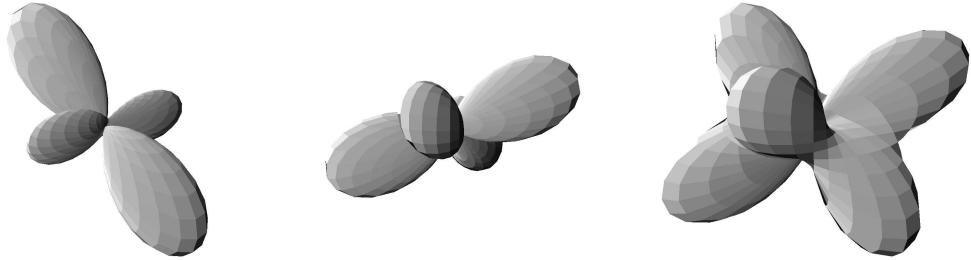}}
\end{displaymath}
\caption{\small Left and middle: real basis functions for the trigonal 
doublet $E_{\rm tr}$. Right: the complex linear combinations have 
unquenched orbital moment along the trigonal $z$ axis.}
\label{fig:trig}
\end{figure}

It is interesting to compare the case of the usual $E$ doublet with the
 doublet $E_{\rm tr}$ which is split off from the $T_2$ level by a trigonal
 distortion, such as seen in the $3d^1$ compound BaVS$_3$ \cite{bavs,pk}. The
 real wave functions are shown in Fig.~\ref{fig:trig} (left and middle), while 
the complex combination on the right. It apparently has a trigonal shape but
 can be shown to carry a remnant of unquenched orbital moment along the
 trigonal $z$-axis. The decomposition for the order parameter 
$E\otimes E=A_1+A_2+E$ is valid here, too, but the basis function of $A_2$ 
is simply $\sim z\rightarrow L^z$, thus time reversal invariance breaking 
can be realized with (orbital) magnetic dipole moment.

It is a great honour to be permitted to dedicate this paper to Martin 
Gutzwiller on the occasion of his 75th birthday.

{\bf Acknowledgements}. The author is greatly indebted to Karlo Penc for
valuable advice and discussions, and for his generous help in many matters. 
Support by the Hungarian research grants  
 OTKA T025505 and AKP 98-66 is gratefully acknowledged.


\begin{thebibliography}{99}  

\bibitem{Gutz1} M.C. Gutzwiller: Phys. Rev. Lett. {\bf 10}, 159 (1963).
\bibitem{Gutz2} M.C. Gutzwiller: Phys. Rev. {\bf 134}, A923 (1964).
\bibitem{Gutz3} M.C. Gutzwiller: Phys. Rev. {\bf 137}, A1726 (1965).
\bibitem{VV} J.H. Van Vleck: Rev. Mod. Phys. {\bf 25}, 220 (1953). 
\bibitem{Hubb} J. Hubbard: Proc. Roy. Soc. London A {\bf 276}, 238 (1963).
\bibitem{Kana} J. Kanamori: Prog. Theor. Phys. {\bf 30}, 275 (1963).
\bibitem{flori} J. B\"unemann, F. Gebhard, and W. Weber: in this volume (see
  also cond-mat/0006238).
\bibitem{scien2000} For a recent brief review, see Y. Tokura and N. Nagaosa:
  Science {\bf 288}, 462 (2000). \cite{LN} gives an elementary treatise of 
the basic facts.  
\bibitem{khom2000} D. Khomskii: cond-mat/0004034.
\bibitem{shiba2000} A. Takahashi and H. Shiba: preprint (2000).
\bibitem{naga2000} R. Maezono and N. Nagaosa: cond-mat/0005482.
\bibitem{MV2} W. Metzner and D. Vollhardt: Phys. Rev. B {\bf 37}, 7382 (1987).
\bibitem{LN} P. Fazekas: {\sl Lecture Notes on Electron Correlation and 
Magnetism}, Series in Modern Condensed Matter Physics Vol. {\bf 5}, 
World Scientific, Singapore (1999).
\bibitem{FMM90} P. Fazekas, B. Menge, E. M\"uller-Hartmann: Z. Phys. B:
  Condens. Matter {\bf 78}, 69 (1990).
\bibitem{MV1} W. Metzner and D. Vollhardt: Phys. Rev. Lett. {\bf 62}, 324
  (1989).
\bibitem{BR70} W.F. Brinkman and T.M. Rice: Phys. Rev. B {\bf 2}, 4302 (1970.
\bibitem{schlipf} J. Schlipf, M. Jarrell, P.G.J. van Dongen, N. Bl\"umer,
     S. Kehrein, D. Vollhardt: Phys. Rev. Lett. {\bf 82}, 4890 (1999).
\bibitem{RU86} T.M. Rice and K. Ueda: Phys. Rev. B {\bf 34}, 6420 (1986).
\bibitem{FB87} P. Fazekas and B.H. Brandow: Phys. Scripta {\bf 36}, 809 (1987).
\bibitem{Geb91} F. Gebhard: Phys. Rev. B{\bf 44}, 992 (1991).
\bibitem{bune} J. B\"unemann, W. Weber, and F. Gebhard: Phys. Rev. B {\bf 57},
  6896 (1998).
\bibitem{itai} K. Itai and P. Fazekas: Phys. Rev. B {\bf 54}, R752 (1996).
\bibitem{penn} D.R. Penn: Phys. Rev. {\bf 142}, 350 (1966).
\bibitem{Kukh} K.I. Kugel and D.I. Khomskii: Sov. Phys. Usp. {\bf 25}, 231
  (1982). 
\bibitem{sakai} O. Sakai, R. Shiina, H. Shiba, P. Thalmeier: 
J. Phys. Soc. Japan {\bf 68}, 1364 (1999).
\bibitem{kura} Y. Kuramoto, H. Kusunose: J. Phys. Soc. Japan {\bf 69}, 671 
(2000).
\bibitem{KK} L.I. Korovin and E.K. Kudinov: Fiz. Tverd. Tela {\bf 16}, 2562
  (1974).
\bibitem{heldvol} K. Held and D. Vollhardt: Phys. Rev. Lett. {\bf 84}, 5168 
(2000).
\bibitem{wahle} J. Wahle, N. Bl\"umer, J. Schlipf, K. Held, and D. Vollhardt:
Phys. Rev. B{\bf 58}, 12749 (1998).
\bibitem{rozenberg} M.J. Rozenberg: Eur. Phys. J. B{\bf 2}, 457 (1998).
\bibitem{pfi} K. Penc, P. Fazekas and K. Itai: in preparation.
\bibitem{bavs} G. Mih\'aly, I. K\'ezsm\'arki, F. Z\'amborszky, M. Miljak,
  K. Penc, P. Fazekas, H. Berger and L. Forr\'o: Phys. Rev. B {\bf 61}, 
R7831 (2000).
\bibitem{pk} K. Penc and P. Fazekas: to be published.
\end{thebibliography}
\end{document}